\documentclass[12pt]{article}
\usepackage{graphicx}
\usepackage{cite}
\usepackage{color}
\def\dlambda{\hspace{-4pt}
\raisebox{2.5pt}{\Large\hspace{3pt} -\hspace{-6.5pt}}\lambda}

\newcommand{\bm}[1]{\mbox{\boldmath $#1$}}
\newcommand{\fnd}[2]{\frac{\textstyle #1}{\textstyle #2}}
\newcommand{\x}[1]{{\textstyle #1}}
\newcommand{\abs}[1]{\left| #1\right|}
\newcommand{\fndrs}[4]{\fnd{\raisebox{#1}{$#2$}}{\raisebox{#3}{$#4$}}}
\newcommand{\xrm}[1]{{\textstyle \mbox{\rm #1}}}
\newcommand{\dissum}[2]{\displaystyle \sum_{#1}^{#2}}
\newcommand{\Real}[1]{\Re {\it e}\left(#1 \right)}
\newcommand{\Imag}[1]{\Im {\it m}\left(#1 \right)}
\newcommand{\bra}[1]{\mbox{$\left\langle #1\right|$}}
\newcommand{\ket}[1]{\mbox{$\left| #1\right\rangle$}}

\newcommand{\braket}[3]{\mbox{$\left\langle #1\left|
#2\right| #3\right\rangle$}}
\begin{document}
\title{Relating multichannel scattering and production amplitudes
in a microscopic OZI-based model}
\author{
Eef van Beveren\\
{\normalsize\it Centro de F\'{\i}sica Te\'{o}rica,
Departamento de F\'{\i}sica, Universidade de Coimbra}\\
{\normalsize\it P-3004-516 Coimbra, Portugal}\\
{\small http://cft.fis.uc.pt/eef}\\ [.3cm]
\and
George Rupp\\
{\normalsize\it Centro de F\'{\i}sica das Interac\c{c}\~{o}es Fundamentais,
Instituto Superior T\'{e}cnico}\\
{\normalsize\it Universidade T\'{e}cnica de Lisboa, Edif\'{\i}cio Ci\^{e}ncia,
P-1049-001 Lisboa, Portugal}\\
{\small george@ist.utl.pt}\\ [.3cm]
{\small PACS number(s): 11.80.Gw, 11.55.Ds, 13.75.Lb, 12.39.Pn}
}
\maketitle

\begin{abstract}
Relations between scattering and production amplitudes are studied
in a microscopic multichannel model for meson-meson scattering, with
coupling to confined quark-antiquark channels. Overlapping resonances
and a proper threshold behaviour are treated exactly in the model.
Under the spectator assumption, it is found that the two-particle
production amplitude shares a common denominator with the elastic
scattering amplitude, besides a numerator consisting of a linear
combination of all elastic and some inelastic matrix elements.
The coefficients in these linear combinations are shown to be generally
complex. Finally, the standard operator expressions relating production
and scattering amplitudes, viz.\ $A=T/V$ and $\Imag{A}=T^{\ast}A$, are
fulfilled, while in the small-coupling limit the usual isobar model is
recovered.
\end{abstract}

\section{Introduction}

In a very recent article \cite{JPG34p1789} we have shown that several
hadronic three-body decays of $J/\psi$, $D$ and $D_s$ mesons can be well
described, up to moderately high energies, in a model for production
processes derived from the so-called Resonance-Spectrum Expansion (RSE)
\cite{IJTPGTNO11p179}. The RSE formalism amounts to an effective
description of non-exotic meson-meson scattering, based upon quark-antiquark
pair creation and annihilation allowing transitions between an infinity of
confined $q\bar{q}$ states and the meson-meson continuum. An essential
feature of the RSE is that it gives rise to closed-form expressions for the
$S$-matrix and even the fully off-shell $T$-matrix. Hence, exact analyticity
and unitarity properties, as well as a correct (sub)threshold behaviour, are
manifestly satisfied. Moreover, the resulting meson-meson production
amplitude can be derived exactly, too, by summing the corresponding two-body
Born series, the only assumption being that the third particle acts as a mere
spectator \cite{JPG34p1789}.

In the present paper, we shall further develop the formalism introduced
in Ref.~\cite{JPG34p1789}, so as to cover the most general multichannel case
in mesonic 3-body decays. Clearly, at higher energies competing inelastic
2-meson channels require that the production amplitude be described by a
vector and not a scalar function. The underlying 2-body scattering $T$-matrix
is then a true matrix. Furthermore, also the quark-antiquark sector needs an
extension, as there can be mixing of different $q\bar{q}$ channels that
couple to the same meson-meson channels. This is the case in e.g.\ the
production of $I\!=\!0$ $\pi\pi$ and $K\bar{K}$ pairs, which both couple to
the $n\bar{n}$ ($=(u\bar{u}+d\bar{d})/\sqrt{2}$) and $s\bar{s}$ channels,
giving rise to the isoscalar scalar resonances $f_0(600)$ (alias $\sigma$)
and $f_0(980)$. Finally, having a general and exact --- within the model
assumptions --- expression for production amplitudes at hand, we may carry
out a detailed comparison with the ansatzes employed in other approaches,
focusing on common features as well as clear differences.

Under the spectator approximation, we assume that a pair of mesons is created
out of one $q\bar{q}$ pair emerging from the original decay, accompanied by a
non-interacting spectator meson. In this process, we only consider OZI-allowed
\cite{OZI} strong transitions to pairs of mesons. The intensities
of transitions between the initial $q\bar{q}$ pair (its quantum numbers,
including flavour, are abbreviated by $\alpha$) and the various meson-meson
pairs $i$ allowed by quantum numbers are given by coupling constants
$g_{\alpha i}$ determined via the recoupling scheme of Ref.~\cite{ZPC21p291}.
In Sec.~\ref{sectTmatrix}, we derive the matrix elements $t(i\to\nu ,E)$
of the scattering amplitude at total CM energy $E=\sqrt{s}$, for transitions
between the meson-meson channels $i$ and $\nu$. In the same section, we
establish a relation between the common denominator $D(E)$ of all matrix
elements $t(\xrm{anything}\to\xrm{anything},E)$ and the numerators of
diagonal matrix elements of $t(E)$,  the latter representing elastic
scattering. The production amplitude $a(\alpha\to i, E)$, which is related to
the probability of producing a meson-meson pair $i$, assuming that a
$q\bar{q}$ pair emerges in the initial --- here not described -- stages of
the decay process, is determined in Sec.~\ref{ProductionMatrix}. Note that
the process giving rise to the initial $q\bar{q}$ pair plus the spectator
meson can be either weak or strong yet OZI-suppressed (see
Ref.~\cite{JPG34p1789} for some examples).

The central result of the present paper is a relation between the production
and scattering amplitudes which can be formulated as
\begin{equation}
a(\alpha\to i, E)\;\propto\;
\fnd{g_{\alpha i}}{D(E)}\, +\,
i\dlambda^{2}\,\sum_{\nu}\,
\left\{\,
g_{\alpha i}\, x_{\nu}(E)\, t(\nu\to\nu ,E)\, -\,
g_{\alpha\nu}\, x_{i}(E)\, t(i\to\nu ,E)
\,\right\}
\; ,
\label{centralresult}
\end{equation}
where the $x_{i}$ stand for momentum distributions that will be specified
in Sec.~\ref{ProductionMatrix} (Eq.~\ref{prodall}).

We thus obtain the result that the production amplitude is in the first
place given by the common denominator of the scattering amplitudes. This
implies that, within the RSE formalism, resonance poles are identical for
production and scattering, at least in the spectator approximation. Secondly,
we find that the remainder of the production amplitude to the $i$-th
two-meson channel is proportional to the sum of the differences between
all possible elastic scattering amplitudes $t(\nu\to\nu ,E)$
and the inelastic amplitudes for the $i$-th channel, $t(i\to\nu ,E)$.
This does not spoil our conclusion about the resonance poles,
since all $T$-matrix elements share the {\it common} \/denominator.
Note, however, that the contribution of the term $\nu=i$ vanishes in the
expression between braces on the r.h.s.\ of Eq.~(\ref{centralresult}).
Consequently, the amplitude $a(\alpha\to i, E)$ for the production of a
two-meson pair $i$ does not carry any dependence on the amplitude for
elastic scattering $i\to i$.  In those cases where the coupling to different
meson-meson channels vanishes or can be neglected, implying a $1\times1$
$T$-matrix, the resulting production amplitude is solely determined by the
common denominator $D(E)$ \cite{JPG34p1789}.

Some words are due  about the $K$-matrix formalism. In general, and so
also here, the $T$-matrix can be written as $T=K/(1-iK)$, where $K$ is
a real symmetric matrix. So, at first sight, it seems that the common
denominator of all $T$-matrix elements is given by $1-iK$. However,
this is not the case. First of all, $K$ is a matrix and not just a real
function. But even in the $1\times 1$ case, where $K$ can be represented
by a real function, it has a denominator itself, the zeros of which
are the $K$-matrix poles, located at the real energies where some
eigenphase shift passes through $90^\circ$. The common denominator above
is then the sum of the denominator of $K$ plus $-i$ times the numerator
of $K$.  In general, when $K$ is a matrix, this relation involves the
determinant of $K$. Hence, comparing $K$-matrix poles, lying on the real
axis in the complex energy plane, and resonance poles, which are usually
in the {\it second} \/Riemann sheet with respect to the nearest ``open''
threshold, is far from trivial. Moreover, for some resonances, like the
$\sigma$ and the $\kappa$ ($K_0^*(800)$ \cite{JPG33p1}), the respective
$K$-matrix poles, corresponding to an elastic phase shift
$\delta=90^\circ$, lie about 350--600 MeV higher in energy that the
real parts of the respective $S$-matrix poles, while mixing with other
resonances ($f_0(980)$ and $K_0^*(1430)$) further complicates the picture.
So rather than making \em ad hoc \em \/assumptions about poles of the
production amplitude, we shall straightforwardly derive the latter, and
then see what its properties are.

A final remark here concerns Watson's \cite{PR88p1163} theorem for
production. This theorem implicitly relies on having a potential
which is energy independent or only weakly energy dependent. However,
this is not the case here, because the energy dependence of our
effective meson-meson potential is far stronger than that of the
scattering $T$-matrix. As a consequence, the energy dependence of the
production experiment does not resemble at all the one of the $T$-matrix,
and all exercises imposing the Watson ``theorem'' or theorems derived
from it are inappropriate here. This issue is analysed in more detail
in Ref.~\cite{NPA744p127}.

\section{The Resonance-Spectrum Expansion (RSE)}

Scattering from a weakly coupled resonant source has been studied in
a variety of different approaches. For such systems it is observed
that resonances occur at energies that are close to the unperturbed
spectrum of the resonant source. Widths and mass shifts can be determined
by perturbative methods, and expressed in terms of pole positions of the
resonances in the complex energy plane.

Intuitively, however,
perturbative methods do not offer the correct
strategy for strong interactions. Since in the present paper we are
interested in obtaining exact relations between scattering and production
amplitudes, which are moreover based on a microscopic description in terms
of quarks, we rather fall back upon an approximate yet exactly solvable
theory or model. Such a manifestly unitary and analytic framework is
provided by the RSE.

The RSE aims at describing the scattering of meson pairs in non-exotic
channels, thereby assuming that in the interaction region a meson pair
may temporarely transform into a quark-antiquark pair through $q\bar{q}$
annihilation and subsequent creation. The transitions of the system,
from meson-meson pairs to $q\bar{q}$ pairs and vice-versa,
are described by an off-diagonal potential $V_{t}$ in the RSE, linking
these two sectors to each other. It has a maximum at an interquark
distance $r_{0}$ which depends on the average effective quark mass and
runs from slightly less than 0.2 fm for $b\bar{b}$ to about 0.6 fm
for light quarks.  Furthermore, we assume that this mechanism gives rise
to the dominant meson-meson interaction in non-exotic channels.
Here, we limit ourselves to the case where $V_{t}$ is considered
the only interaction.

The intermediate $q\bar{q}$ states are supposed to have an unperturbed
confinement spectrum depending on the quantum numbers of the system.
Its energy eigenvalues are all contained in the two-meson scattering
matrix, which, consequently, develops corresponding CDD resonance poles.

An addtional nice feature of the RSE, which will turn out to be crucial
for the construction of the production amplitude, is the possibility to
obtain the closed-form scattering $T$-matrix both in configuration and
in momentum space. In the former representation, coupled-channel
Schr\"{o}dinger equations with the proper boundary conditions directly
lead to the full solution, while in the latter picture individual Born
terms can be explicitly calculated and then exactly summed up owing to
the general separable nature of the effective meson-meson potential,
with no need to solve the Lippmann-Schwinger \cite{PR79p469} integral
equations. This allows to verify the correctness of the momentum-space
approach in the scattering case, which is the only method at our disposal
to describe production. As we shall see below, a similar Born series can
then be written down and summed up.

\section{Scattering}
\label{sectTmatrix}

\begin{figure}[htbp]
\begin{center}
\begin{tabular}{cc}
\resizebox{!}{60pt}{\includegraphics{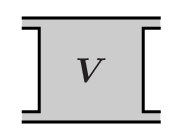}} &
\resizebox{!}{60pt}{\includegraphics{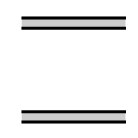}}\\
(effective 2-meson potential) & (two-meson propagator)
\\ [-15pt]
\end{tabular}
\end{center}
\caption[]{\small
Graphical representation of the building blocks of the RSE two-meson
scattering amplitude. The solid lines represent valence quarks and
antiquarks as in usual Feynman diagrams. In contrast, the gray areas
stand for all possible confining interactions, like gluon exchange,
sea-quark loops and their higher orders. The effective meson-meson
interaction is represented by $V$. Furthermore, although the mesons
in the two-meson propagators are considered pointlike in the RSE,
for clarity they are here represented by double quark lines connected
by confining interactions.}
\label{loops}
\end{figure}

The building blocks of the RSE meson-meson scattering amplitude
are the effective meson-meson potentials and the free two-meson
propagators, which are graphically represented in Fig.~\ref{loops}.
It is the philosophy of the RSE that confinement and decay can be separated.
Hence, in the interaction region, a two-meson system can appear
as a permanently confined system consisting of a valence quark and a valence
antiquark. Possible intermediate crypto-exotic multiquark states are not
considered in the RSE.

\begin{figure}[htbp]
\begin{center}
\begin{tabular}{c}
\resizebox{0.94\textwidth}{!}{\includegraphics{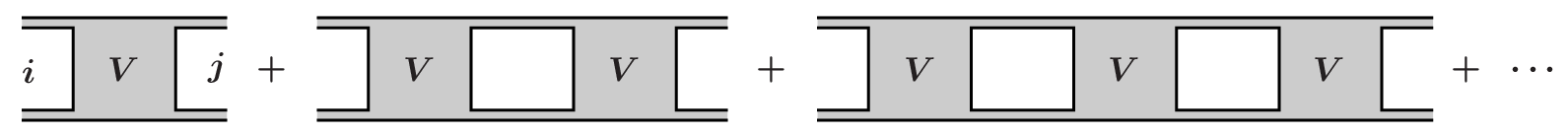}}
\\ [-20pt]
\end{tabular}
\end{center}
\caption[]{
Graphical representation of the RSE scattering amplitude.}
\label{theTmatrix}
\end{figure}

The dynamics of the intermediate $q\bar{q}$ states is described by a
permanently confining Hamiltonian $H_{c}$, which has a complete set of
eigenstates at eigenvalues representing the confinement spectrum.
The other part of the strong interactions, generating transitions between
a two-meson system and a $q\bar{q}$ state, is given by a transition potential
$V_{t}$.  Consequently, the effective meson-meson interaction in
Fig.~\ref{loops} is described by the operator
\begin{equation}
V\; =\;{V_{t}}^{T}\;\left[ E-H_{c}\right]^{-1}\; V_{t}
\;\;\; ,
\label{quarkloop}
\end{equation}
where $E$ is the total invariant mass of the coupled-channel system.
These interactions and the free two-meson propagators, both depicted
in Fig.~\ref{loops}, can then be used in an ordinary
Lippmann-Schwinger \cite{PR79p469} approach to scattering. Nevertheless,
in constructing the Born series, quark-loop contributions to all orders
are automatically accounted for, as becomes clear from Fig.~\ref{theTmatrix}.

The RSE scattering amplitude depicted in this figure has the Born term
\begin{equation}
V(i\to j)\; =\;\bra{i,\,\vec{p}_{i}\,}\;{V_{t}}^{T}\;
\left[ E(p)-H_{c}\right]^{-1}\; V_{t}\;
\ket{j,\,\vec{p}_{j}\,}
\;\;\; .
\label{defBornterm}
\end{equation}
In Ref.~\cite{IJTPGTNO11p179} it was shown how, under the RSE assumptions,
the integrations can be done analytically.
For the present discussion, it is only necessary to mention the
generic form of the RSE expression, given by
\begin{equation}
V(i\to j)\; =\;\fnd{\dlambda^{2}}{4\pi^{2}}\,
\sum_{\ell =0}^{\infty}(2\ell +1)\,
P_{\ell}\left(\hat{p}_{i}\cdot\hat{p}_{j}\,\right)\,
j_{\ell}(p_{i}r_{0})\, j_{\ell}(p_{j}r_{0})\,
Z^{(\ell )}_{ij}(E)
\;\;\; .
\label{Bornterm}
\end{equation}
The overall coupling $\dlambda$ and the interaction radius $r_{0}$
represent the total probability of quark-pair creation/annihilation
and the average interquark distance at which such processes take place,
respectively; $j_{\ell}$ stands for the spherical Bessel function
for the relative meson-meson angular momentum $\ell$;
$\vec{p}_{i}$ and $\vec{p}_{j}$ are the relative linear momenta in
the two-meson channels $i$ and $j$, respectively.
The matrix $Z$ is a real and symmetric function of the total invariant
mass of the system.

The intermediate $q\bar{q}$ systems may have different orbital angular
momenta, for the same quantum numbers. For example, a meson-meson system
with $J^{PC}=1^{--}$ couples to $q\bar{q}$ systems in either an $S$ or a
$D$ wave. On the other hand, isosinglet mesons can be mixtures of
different quark flavours, usually $n\bar{n}$ and $s\bar{s}$.
In such cases, more than one type of $q\bar{q}$ states are involved
in the quark loops of the process depicted in Fig.~\ref{theTmatrix}.
However, since the recurrencies of the permanently confined
$q\bar{q}$ systems are automatically included by the definition of the
Born term (\ref{Bornterm}), the number of $q\bar{q}$ channels that couple
to a specific set of two-meson quantum numbers is limited, usually to one
or two. Nevertheless, the number of two-meson scattering channels is in
principle not restricted.

For any number of coupled confinement and scattering channels,
the general structure of the amplitude reads
($E=E(p_{i})=E(p_{j})$)
\begin{eqnarray}
\lefteqn{
t(i\to j)\, =\,\braket{i,\,\vec{p}_{i}\,}{t}{j,\,\vec{p}_{j}\,}
\, =\,
\braket{i,\,\vec{p}_{i}\,}{(V+VGV+VGVGV+\dots )}{j,\,\vec{p}_{j}\,}
}
\label{tmatrix}\\ [10pt] & & =\,
\fnd{\dlambda^{2}}{4\pi^{2}}\,
\sum_{\ell =0}^{\infty}(2\ell +1)\,
P_{\ell}\left(\hat{p}_{i}\cdot\hat{p}_{j}\,\right)\,
j_{\ell}\left( p_{i}r_{0}\right)\, j_{\ell}\left( p_{j}r_{0}\right)\;
\fnd{{\cal A\,}^{(\ell )}_{ij}(E)}{{\cal D\,}^{(\ell)}(E)}
\;\;\; ,
\;\;\;\;\;\;\;\;\;\;
\;\;\;\;\;\;\;\;\;\;
\nonumber
\end{eqnarray}
where $\cal A$ and $\cal D$ are functions of
the total invariant mass $E$
satisfying the unitarity condition
\begin{equation}
\Imag{{\cal D\,}^{(\ell)}{{\cal A\,}^{(\ell )}_{ij}}^{\ast}}
\, =\,
2\dlambda^{2}\,
\sum_{\nu}\,
\mu_{\nu}p_{\nu}j_{\ell}^{2}\left( p_{\nu}r_{0}\right)\,
{\cal A\,}^{(\ell)}_{i\nu}{{\cal A\,}^{(\ell)}_{j\nu}}^{\ast}
\;\;\; .
\label{UnitaircompleteTell}
\end{equation}

The denominator $\cal D$ contains the full pole structure of
the coupled two-meson states. In order to be a bit more specific,
let us consider the scattering of charmed mesons, i.e.,
$D\bar{D}$, $D^{\ast}\bar{D}$, $D^{\ast}\bar{D}^{\ast}$,
$D_{s}\bar{D}_{s}$, $D^{\ast}_{s}\bar{D}_{s}$
and $D^{\ast}_{s}\bar{D}^{\ast}_{s}$,
all coupled to $c\bar{c}$.
For such a process, $\cal D$ has in the RSE the form
\begin{equation}
{\cal D}^{(\ell )}(E)\; =\;
1+2i\dlambda^{2}
\sum_{\nu}
g_{\nu}^{2}
\left\{\dissum{n=0}{\infty}
\fnd{\abs{F_{c\bar{c}}^{(n)}\left( r_{0}\right)}^{2}}{E-E_{n}}\right\}
\mu_{\nu}p_{\nu}
j_{\ell}\left( p_{\nu}r_{0}\right)
h^{(1)}_{\ell}\left( p_{\nu}r_{0}\right)
\;\;\; ,
\label{ccbarDD}
\end{equation}
where the outer sum runs over all two-meson channels,
and the inner sum over all recurrencies $n$ for the operator $H_{c}$
describing confinement in the $c\bar{c}$ system.
$F_{c\bar{c}}^{(n)}$ and $E_{n}$ represent
the eigenstate and eigenvalue of the $n$-th
recurrency of the $H_{c}$ spectrum, respectively.
Furthermore, the $g_{\nu}$ stand for the relative couplings of each of
the two-meson systems to $c\bar{c}$, while
$h^{(1)}_{\ell}$ is a spherical Hankel function of the first kind.

The denominator ${\cal D}(E)$ vanishes for $E$ near $E_{n}$ and
small overall coupling $\dlambda$. In this case, the scattering cross
sections in all channels display narrow spikes for values of $E$
in the vicinity of $E_{n}$ ($n=0$, 1, 2, $\dots$).
Hence, for small $\dlambda$, the theoretical cross sections reproduce
--- up to small shifts ---  the hypothetical $c\bar{c}$ confinement
spectrum.

However, for larger values of $\dlambda$ the zeros in $\cal D$
are no longer near the eigenvalues of $H_{c}$, but move deeper into
the complex $E$ plane, farther away from the real axis
and with appreciable shifts for the real parts as well.
Then, the resonance spectrum does no longer reproduce the spectrum of
$H_{c}$: resonances start overlapping and even the number of zeros in
$\cal D$ that lie close enough to the real energy axis to be observed
experimentally may change. We believe this describes quite accurately
the true situation in hadron spectroscopy.

Below the lowest threshold, poles,
i.e., zeros in $\cal D$ (Eq.~\ref{ccbarDD}),
come out on the real axis, because the expression
$ij_{\ell}h^{(1)}_{\ell}$ turns real.
In that case, expression~(\ref{tmatrix}) describes bound $c\bar{c}$
states, such as $\eta_{c}$, $J/\psi$, $\chi_{c}(1P)$ and $\psi (2S)$,
yet with an admixture of two-meson components.
The energy eigenvalues of these ``dressed'' states then depend on the
value of $\dlambda$. It has been observed \cite{PRD21p772,AIPCP814p143}
that charmonium mass shifts with respect to the pure confinement spectrum
can be surprisingly large in the RSE, as well as in other approaches
\cite{PRD72p034010}.

In the present work, we intend to derive relations among
${\cal A\,}^{(\ell )}_{ij}$,
${\cal D\,}^{(\ell)}$ and $Z^{(\ell )}_{ij}$.
In principle, this could be achieved by just performing the calculus
outlined in Ref.~\cite{IJTPGTNO11p179}. However, here we shall allow
more general expressions for the $Z$ matrix in the Born term (\ref{Bornterm}).
Hence, apart from the unitarity condition~(\ref{UnitaircompleteTell}),
we must construct a second relation. For that purpose, we write the identity
\begin{eqnarray}
\lefteqn{0\, =\,
\braket{i,\,\vec{p}_{i}\,}{(T-V-TGV)}{j,\,\vec{p}_{j}\,}
\, =\,
\fnd{\dlambda^{2}}{4\pi^{2}}\,
\sum_{\ell =0}^{\infty}(2\ell +1)\,
P_{\ell}\left(\hat{p}_{i}\cdot\hat{p}_{j}\,\right)\,
j_{\ell}(p_{i}r_{0})\, j_{\ell}(p_{j}r_{0})
\,\times}
\label{TGVgenN}\\ [10pt] & &
\times\,\left\{
\fnd{{\cal A}^{(\ell )}_{ij}(E)}{{\cal D}^{(\ell )}(E)}
\, -\,
Z^{(\ell )}_{ij}(E)
\, +\, 2i\dlambda^{2}
\sum_{\nu}\,\mu_{\nu}p_{\nu}
j_{\ell}\left( p_{\nu}r_{0}\right)\,
h^{(1)}_{\ell}\left( p_{\nu}r_{0}\right)\,
\fnd{{\cal A}^{(\ell )}_{i\nu}(E)}{{\cal D}^{(\ell )}(E)}
Z^{(\ell )}_{\nu j}(E)
\right\}
\;\;\; ,
\;\;\;\;\;\;\;\;\;\;
\nonumber
\end{eqnarray}
which yields the relation
\begin{equation}
{\cal D}^{(\ell )}\, Z^{(\ell )}_{ij}\; =\;
{\cal A}^{(\ell )}_{ij}
\, +\, 2i\dlambda^{2}
\sum_{\nu}\,\mu_{\nu}p_{\nu}
j_{\ell}\left( p_{\nu}r_{0}\right)\,
h^{(1)}_{\ell}\left( p_{\nu}r_{0}\right)\,
{\cal A}^{(\ell )}_{i\nu}
Z^{(\ell )}_{\nu j}
\;\;\; .
\label{ABeqApsum}
\end{equation}

Furthermore, if we assume
\begin{equation}
{\cal A}^{(\ell )}_{ij}\; =\;
{\cal A}^{(\ell )(0)}_{ij}
\, +\,\dlambda^{2}\,{\cal A}^{(\ell )(1)}_{ij}
\, +\,\dlambda^{4}\,{\cal A}^{(\ell )(2)}_{ij}
\, +\,\dots
\;\;\; ,
\label{calAansatz}
\end{equation}
then we obtain the following solution to relations
(\ref{UnitaircompleteTell}) and (\ref{ABeqApsum}):
\begin{enumerate}
\item
The denominator $\cal D$ can be fully expressed in terms
of the numerators $\cal A$, according to
\begin{equation}
{\cal D}^{(\ell )}\; =\;
1\, +\, 2i\dlambda^{2}\,\sum_{\nu}\,
\mu_{\nu}\, p_{\nu}\,
j_{\ell}\left( p_{\nu}r_{0}\right)\,
h^{(1)}_{\ell}\left( p_{\nu}r_{0}\right)\,
{\cal A}^{(\ell )}_{\nu\nu}
\;\;\; .
\label{solBansatz}
\end{equation}
\item
The zeroth- order term of (\ref{calAansatz}) is evidently
given by the Born term (\ref{Bornterm}):
\begin{equation}
{\cal A}^{(\ell )(0)}_{ij}\; =\;
Z^{(\ell )}_{ij}
\;\;\; .
\label{sol0Aansatz}
\end{equation}
\item
For the higher-order terms of the expansion (\ref{calAansatz})
we obtain the recursion relation
\begin{equation}
{\cal A}^{(\ell )(n+1)}_{ij}\; =\;
2i\,\sum_{\nu}\,
\mu_{\nu}\, p_{\nu}\,
j_{\ell}\left( p_{\nu}r_{0}\right)\,
h^{(1)}_{\ell}\left( p_{\nu}r_{0}\right)\,
\left\{
{\cal A}^{(\ell )(n)}_{\nu\nu}\, Z^{(\ell )}_{ij}
\, -\,
{\cal A}^{(\ell )(n)}_{i\nu}\, Z^{(\ell )}_{\nu j}
\right\}
\; .
\label{solAansatz}
\end{equation}
\end{enumerate}

From Eq.~(\ref{tmatrix}) we then get a partial-wave scattering amplitude
of the form
\begin{equation}
t_{\ell}(i\to j)\; =\;
2\dlambda^{2}\,
j_{\ell}\left( p_{i}r_{0}\right)\, j_{\ell}\left( p_{j}r_{0}\right)\;
\fnd{{\cal A\,}^{(\ell )}_{ij}(E)}{{\cal D\,}^{(\ell)}(E)}
\;\;\; .
\label{tpwamplitude}
\end{equation}
For a full definition of this amplitude, satisfying the unitarity conditions
for scattering, see Eq.~(\ref{Tpartialwave}).

\section{Production}
\label{ProductionMatrix}

Various opinions exist on how to analyse the final-state interactions
of pairs of hadrons emerging from a decay process
\cite{PAN67p614,PLB585p200,HEPPH0703256,ARXIV07061341}.
In particular, the production of pion pairs has been studied from many
different angles. Several resonances have been discovered and established
in this channel. However, there still are many open questions, of which the
most intriguing one probably is the formation of the $f_{0}(980)$ resonance
\cite{PLB495p300,PLB521p15,AIPCP619p112,PLB527p193,PRD67p014012,PLB559p49,
PRD68p036001,PLB586p53,PAN68p1554,EPJA24p437,EPJC47p45,HEPPH0606266,
PRD74p114001}.
As such, this resonance seems to be one of the key issues
for understanding strong interactions. It lies close the $K\bar{K}$
threshold, couples relatively weakly to pions, comes on top of a much broader
structure, namely the $f_{0}(600)$, and is furthermore not very distant
from a broad resonance around 1.35 GeV, viz.\ the $f_{0}(1370)$
\cite{ARXIV07061341}.

It is our understanding that mesonic resonances, like the $f_{0}(600)$ and
the $f_{0}(980)$, form an integral part of the whole meson family.
Therefore, we have developed a model for all $q\bar{q}$ phenomena,
including those involving charm and bottom.
Here, we wish to develop a new tool for data analysis,
which is an amplitude for the description of final-state interactions
in two-meson subsystems emerging in decay processes involving other
particles. This production amplitude is based on the two-meson scattering
amplitude given in Eq.~(\ref{tmatrix}).
\begin{figure}[htbp]
\begin{center}
\begin{tabular}{c}
\resizebox{0.94\textwidth}{!}{\includegraphics{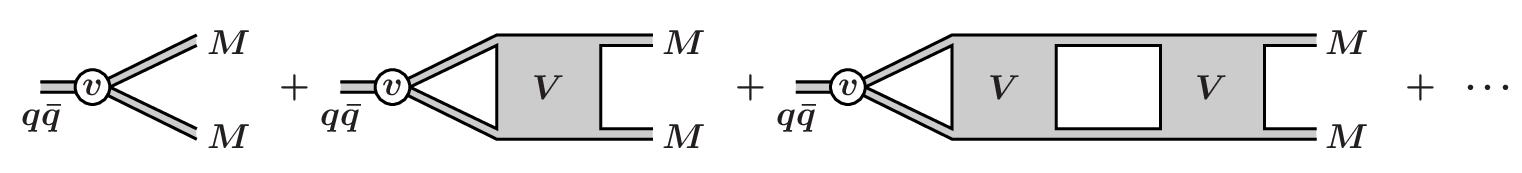}}
\\ [-20pt]
\end{tabular}
\end{center}
\caption[]{\small
Graphical representation of the RSE production amplitude.
The transition $q\bar{q}\to MM$, denoted by $V_{t}$ in the text,
is here represented by $v$; the resulting effective $MM$ interaction
is denoted by $V$.}
\label{theAmatrix}
\end{figure}

For the description of the final-state interactions of meson pairs
in production processes, it is common practice to make the spectator
assumption, according to which the other emerging hadrons do not
interact strongly with the pair. Evidently, this is an approximation,
which is justified by the observation that in most production processes
resonances involving the third (or fourth, \ldots) hadron are much higher
in mass than the energies considered for the pair.
Here, we moreover assume that the meson pair is generated from
an initially produced $q\bar{q}$ pair. Our amplitude for the production
of a meson pair, including all higher-order contributions from final-state
interactions, is depicted in Fig.~\ref{theAmatrix}.
Also using expression~(\ref{tmatrix}) for the scattering amplitude,
we are led to define for the production amplitude
\begin{eqnarray}
\lefteqn{a(\alpha\to i)\, =\,
\braket{i,\,\vec{p}_{i}}{(1+TG)V_{t}}
{\left( q\bar{q}\right)_{\alpha}\, ,\, E}
\, =}
\label{prodall_MxN}\\ [10pt] & & =\,
\braket{i,\,\vec{p}_{i}}{V_{t}}
{\left( q\bar{q}\right)_{\alpha}\, ,\, E}
\, +\,
\sum_{\nu}\int d^{3}k_{\nu}\,
\braket{i,\,\vec{p}_{i}}{T}{\nu ,\,\vec{k}_{\nu}}
G\left(\vec{k}_{\nu}\right)
\braket{\nu ,\,\vec{k}_{\nu}}{V_{t}}
{\left( q\bar{q}\right)_{\alpha}\, ,\, E}
\nonumber\\ [10pt] & & =\,
\fnd{\dlambda}{\sqrt{\pi}}\,
\sum_{\ell ,m}\, (-i)^{\ell}\, j_{\ell}\left( p_{i}r_{0}\right)\,
Y^{(\ell )}_{m}\left(\,\hat{p}_{i}\,\right)\,
Q^{(\alpha )}_{\ell_{q\bar{q}}}\left( E\right)\,
\times
\nonumber\\ [10pt] & &
\;\;\;\;\;\;\;\;\;
\times\,
\left\{ g_{\alpha i}
\, -\,
2i\dlambda^{2}\;
\sum_{\nu}\,
\mu_{\nu}\, p_{\nu}\,
j_{\ell}\left( p_{\nu}r_{0}\right)\,
h^{(1)}_{\ell}\left( p_{\nu}r_{0}\right)\,
g_{\alpha\nu}\,
\fnd{{\cal A\,}^{(\ell )}_{i\nu}(E)}{{\cal D\,}^{(\ell )}(E)}\,
\right\}
\;\;\; .
\nonumber
\end{eqnarray}
Here, $Q^{(\alpha )}_{\ell_{q\bar{q}}}$ represents the overlap with
the initial $q\bar{q}$ distribution, having quantum numbers $\alpha$
and relative interquark angular momentum $\ell_{q\bar{q}}$.
Notice that the latter quantum number is related - though unequal -
to the relative two-meson angular momentum $\ell$,
because of total-angular-momentum and parity conservation.
Below, we shall discuss the properties of production
amplitude~(\ref{prodall_MxN}) for pairs of interacting mesons.

\subsection{\bm{P_{i}=\dissum{\nu}{}c_{\nu}T_{\nu i}}?}

The result (\ref{prodall_MxN}) agrees to some extent
with the expression proposed in Refs.~\cite{DAP507p404,PRD35p1633}.
Like here, the authors of Ref.~\cite{PRD35p1633} based their ansatz
on the OZI rule \cite{OZI} and the spectator picture, so as
to find that the production amplitude
can be written as a linear combination of the elastic $t_{\ell}(i\to i)$
and inelastic $t_{\ell}(i\to\nu\neq i)$ scattering amplitudes,
with coefficients that do not carry any singularities,
but are rather supposed to depend smoothly on
the total CM energy of the system.

Indeed, if we carry out the substitution (\ref{tpwamplitude}),
we find for our production amplitude the expression
\begin{equation}
a(\alpha\to i)=
\fnd{\dlambda}{\sqrt{\pi}}
\sum_{\ell ,m}\, (-i)^{\ell}
Y^{(\ell )}_{m}\left(\,\hat{p}_{i}\,\right)
Q^{(\alpha )}_{\ell_{q\bar{q}}}\left( E\right)
\left\{ g_{\alpha i}\, j_{\ell}\left( p_{i}r_{0}\right)
-i\sum_{\nu}\,
\mu_{\nu}\, p_{\nu}\,
h^{(1)}_{\ell}\left( p_{\nu}r_{0}\right)\,
g_{\alpha\nu}\,
t_{\ell}(i\to\nu)
\right\} \; ,
\label{prodallint}
\end{equation}
which contains a linear combination of elements of
the scattering amplitude, with coefficients smooth in $E$.

However, Ref.~\cite{PRD35p1633} concluded from the relation
\begin{equation}
\Imag{A}\; =\; T^{\ast}\, A
\label{ImaTstera}
\end{equation}
that the production amplitude must be given
by a \em real \em \/linear combination of the elements of the
transition matrix. A similar conclusion, based on a
$K$-matrix parametrisation, can be found in Ref.~\cite{EPJC30p503}.
In contrast, we arrive at a different conclusion,
namely that, as the Hankel function of the first kind
is a complex function for real arguments,
the coefficients must be {\em complex\em}, in agreement with experimental
analyses \cite{PLB585p200,IJMPA20p482,ARXIV07052248} as well as with the
theoretical work of the Ishidas
\cite{PTP99p1031,AIPCP688p18}.

Relation (\ref{ImaTstera}), which can be also found in
Ref.~\cite{NPA679p671} basically stems from the operator relations
$AV=(1+TG)V=V+TGV=T$, the symmetry of $T$, the realness of $V$
and the unitarity of $1+2iT$,
which gives $\Imag{A}V=\Imag{AV}=\Imag{T}=T^{\ast}T=T^{\ast}AV$.
This leads, for non-singular potentials $V$,
to relation~(\ref{ImaTstera}).
In Appendix~\ref{PvsS}, we show that
notwithstanding the \em complex \em \/coefficients
in Eq.~(\ref{prodallint}),
relation~(\ref{ImaTstera}) is satisfied for the
scattering and production amplitudes
of Eqs. (\ref{tmatrix}) and (\ref{prodall_MxN}), respectively.
Consequently, relation~(\ref{ImaTstera}) does not impose
a realness condition on the coefficients in Eq.~(\ref{prodallint}).

\subsection{The lowest-order term}

Besides the sum over transition matrix elements, our prodution
amplitude~(\ref{prodallint}) also contains an extra term
$\propto g_{\alpha i}\, j_{\ell}\left( p_{i}r_{0}\right)$.
Such a term was not considered in
Refs.~\cite{DAP507p404,PRD35p1633,EPJC30p503}.
However, in the works of Graves-Morris \cite{NCA50p681}
and Aitchison \& collaborators \cite{NPB97p227,JPG3p1503,PLB84p349},
the possible existence of an additional real contribution was anticipated.
Here, it follows straightforwardly from the reasonable assumption that
the produced meson pair originates from an initial $q\bar{q}$ pair.

It is generally agreed that production and scattering
have the same singularity structure in the complex energy plane.
At first sight, this is not obvious from expressions~(\ref{prodall_MxN})
and (\ref{prodallint}). However, the second term between braces in
Eq.~(\ref{prodall_MxN}) can, using Eq.~(\ref{solBansatz}),
be rewritten as follows:
\begin{eqnarray}
\lefteqn{
g_{\alpha i}
\, -\,
2i\dlambda^{2}\;
\sum_{\nu}\,
\mu_{\nu}\, p_{\nu}\,
j_{\ell}\left( p_{\nu}r_{0}\right)\,
h^{(1)}_{\ell}\left( p_{\nu}r_{0}\right)\,
g_{\alpha\nu}\,
\fnd{{\cal A\,}^{(\ell )}_{i\nu}}{{\cal D\,}^{(\ell )}}\,
=}
\label{prodpartly}\\ [10pt] & & =\,
\fnd{1}{{\cal D\,}^{(\ell )}}\,
\left\{
g_{\alpha i}
\, +\,
2i\dlambda^{2}\;
\sum_{\nu}\,
\mu_{\nu}\, p_{\nu}\,
j_{\ell}\left( p_{\nu}r_{0}\right)\,
h^{(1)}_{\ell}\left( p_{\nu}r_{0}\right)\,
\left[\,
g_{\alpha i}\,{\cal A}^{(\ell )}_{\nu\nu}\,
-\,
g_{\alpha\nu}\,{\cal A\,}^{(\ell )}_{i\nu}\,
\right]
\right\}
\nonumber\\ [10pt] & & =\,
\fnd{g_{\alpha i}}{{\cal D\,}^{(\ell )}}\,
\, +\,
2i\dlambda^{2}\;
\sum_{\nu\neq i}\,
\mu_{\nu}\, p_{\nu}\,
j_{\ell}\left( p_{\nu}r_{0}\right)\,
h^{(1)}_{\ell}\left( p_{\nu}r_{0}\right)\,
\left[\,
g_{\alpha i}\,
\fnd{{\cal A}^{(\ell )}_{\nu\nu}}{{\cal D\,}^{(\ell )}}\,
-\,
g_{\alpha\nu}\,
\fnd{{\cal A\,}^{(\ell )}_{i\nu}}{{\cal D\,}^{(\ell )}}\,
\right]
\; .
\nonumber
\end{eqnarray}
From this equation it is obvious that, in our approach,
scattering and production have exactly
the same poles in the complex energy plane,
as they share the global denominator $\cal D$.

\subsection{The central result}

The pole structure of our production amplitude
is exhibited very explicitly in formula~(\ref{prodpartly}),
and shows that it is completely given by $\cal D$,
the very same denominator that determines the pole structure
for elastic scattering. The conclusion is that
resonance shapes are different for production and scattering
because they are largely determined by the respective
\em numerators. \em Moreover,
precisely the numerator ${\cal A}_{ii}$
describing elastic scattering in the $i$-th two-meson channel
has dropped out of expression~(\ref{prodpartly}).
Hence, when restricted to a one-channel model,
our production amplitude is completely determined
by just the denominator $\cal D$.

The result~(\ref{prodpartly}) may be substituted into
relation (\ref{prodall_MxN}).
Moreover, using expression~(\ref{tpwamplitude})
for the partial-wave amplitudes, we arrive at
\begin{eqnarray}
\lefteqn{
a(\alpha\to i)
\, =\,
\fnd{\dlambda}{\sqrt{\pi}}\,
\sum_{\ell ,m}\, (-i)^{\ell}\, j_{\ell}\left( p_{i}r_{0}\right)\,
Y^{(\ell )}_{m}\left(\,\hat{p}_{i}\,\right)\,
Q^{(\alpha )}_{\ell_{q\bar{q}}}\left( E\right)\,
\times}
\nonumber\\ [10pt] & &
\;\;\;\;\;\;\;\;\;
\times\,
\left\{
\fnd{g_{\alpha i}}{{\cal D\,}^{(\ell )}}\,
+\,
i\sum_{\nu\neq i}\,
\mu_{\nu}\, p_{\nu}\,
h^{(1)}_{\ell}\left( p_{\nu}r_{0}\right)\,
\left[\,
g_{\alpha i}\,
\fnd{t_{\ell}(\nu\to\nu)}{j_{\ell}\left( p_{\nu}r_{0}\right)}\,
-\,
g_{\alpha\nu}\,
\fnd{t_{\ell}(i\to\nu)}{j_{\ell}\left( p_{i}r_{0}\right)}\,
\right]
\right\}
\;\;\; .
\label{prodall}
\end{eqnarray}

Equation~(\ref{prodall}) is the central result of our paper.
It explicitly relates the ingredients of elastic scattering
to the amplitude for production in the spectator approximation.
We were able to achieve this because in the RSE one can determine
in an analytically closed form all terms of the perturbation expansions
(\ref{tmatrix}) \cite{IJTPGTNO11p179} and (\ref{prodall_MxN}).
Hence, relations
(\ref{solBansatz}), (\ref{sol0Aansatz}) and (\ref{solAansatz})
can be derived and explicitly verified.
We may thus conclude that at least for a nonrelativistic (NR) microscopic
model, i.e., at low energies, production and scattering are related to
one another through Eq.~(\ref{prodall}).

\subsection{\bm{P=T/V}}

Expression (\ref{prodpartly}) takes an extremely simple form
in the case that all inelasticity is either absent or neglected.
For the $\ell$-th partial wave of
the production amplitude (\ref{prodall_MxN}), we then obtain
\begin{equation}
a^{(\ell )}\;\propto\;
\dlambda\,
j_{\ell}\left( pr_{0}\right)\,
Q^{(\alpha )}_{\ell_{q\bar{q}}}\left( E\right)\,
\fnd{1}{{\cal D\,}^{(\ell )}}\,
\;\;\; .
\label{Ppartialwave}
\end{equation}
This is exactly the generic form of the production amplitude
used in a paper by Roca, Palomar, Oset and Chiang
\cite{NPA744p127},
when at the $f_{0}$(600) resonance
the inelastic contribution $KK\to\pi\pi$ is neglected,
resulting in $a\propto T/V$.
Here, we get in the 1-channel case from Eq.~(\ref{sol0Aansatz})
that ${\cal A}=Z$, which then precisely yields $T/V=1/{\cal D}$.

\subsection{The meson-loop phase}

In the one-channel approximation, we obtain
from the scattering amplitude (\ref{tmatrix})
for the cotangent of the scattering phase shift $\delta^{(\ell )}(E)$
the expression
\begin{equation}
\xrm{cotg}\left(\delta^{(\ell )}(E)\right)
\; =\;
\fndrs{3pt}{n_{\ell}\left( pr_{0}\right)}
{-3pt}{j_{\ell}\left( pr_{0}\right)}\; -\;
\fndrs{0pt}{1}
{-5pt}{2\dlambda^{2}\mu p\,
j_{\ell}^{2}\left( pr_{0}\right)\,
{\cal A\,}^{(\ell )}}
\;\;\; ,
\label{cotangentd}
\end{equation}
where the spherical Neuman function is represented by $n_{\ell}$.

Now, $\cal D$ in formule (\ref{Ppartialwave})
is related to $\cal A$ in formula (\ref{cotangentd})
through Eq.~(\ref{solBansatz}).
After some algebra, we get
\begin{equation}
a^{(\ell )}\;\propto\;
\dlambda\,
j_{\ell}\left( pr_{0}\right)\,
Q^{(\alpha )}_{\ell_{q\bar{q}}}\left( E\right)\,
\left\{
1\, -\,\fnd{\tan\left(\delta^{(\ell )}(E)\right)}
{j_{\ell}\left( pr_{0}\right)/n_{\ell}\left( pr_{0}\right)}
\right\}\,
\cos\left(\delta^{(\ell )}(E)\right)\,
e^\x{i\delta^{(\ell )}(E)}
\;\;\; .
\label{Prodindelta}
\end{equation}

For $S$-waves ($\ell =0$) this becomes
\begin{equation}
a^{(0)}\;\propto\;
\dlambda\,
j_{0}\left( pr_{0}\right)\,
Q^{(\alpha )}_{\ell_{q\bar{q}}}\left( E\right)\,
\left\{
1\, +\,\fnd{\tan\left(\delta^{(0)}(E)\right)}
{\tan\left( pr_{0}\right)}
\right\}\,
\cos\left(\delta^{(0)}(E)\right)\,
e^\x{i\delta^{(0)}(E)}
\;\;\; .
\label{Prodindelta0}
\end{equation}
With respect to the dependence on the phase $\delta^{(0)}(E)$,
this expression has exactly the same form as the $S$-wave production
amplitude given by Boito and Robilotta in Ref.~\cite{ARXIV07053260},
which is based on Watson's formalism \cite{PR88p1163}
via the work of Pennington \cite{HEPPH9710456}.
For the meson-loop phase $\omega (s)$
defined in Ref.~\cite{ARXIV07053260},
we obtain here $pr_{0}$.
However, our resonance poles are determined
in quite a different manner than in
Ref.~\cite{ARXIV07053260}.
Whereas in the RSE the resonance poles are all contained
in $\cal A$ in expression~(\ref{cotangentd})
for the cotangent of the phase shift,
in the formalism employed in Ref.~\cite{ARXIV07053260}
each of the resonance poles for $S$-wave production
has to be put into the corresponding expression by hand, one by one.

\subsection{Breit-Wigner resonances}

Again in the one-channel case, one deduces from Eq.~(\ref{ccbarDD})
for $\cal D$ the form
\begin{equation}
{\cal D}^{(\ell )}(E)\; =\;
1+2i\dlambda^{2}
\left\{\dissum{n=0}{\infty}
\fnd{\abs{F^{(n)}\left( r_{0}\right)}^{2}}{E-E_{n}}\right\}
\mu p\,
j_{\ell}\left( pr_{0}\right)
h^{(1)}_{\ell}\left( pr_{0}\right)
\;\;\; .
\label{D1channel}
\end{equation}
For small $\dlambda$ one finds a zero of $\cal D$
in the vicinity of $E_{n}$, say at $E_{n}\, +\,\Delta E_{n}$,
where
\begin{equation}
\Delta E_{n}\,\approx\,
2\dlambda^{2}\,\abs{F^{(n)}\left( r_{0}\right)}^{2}
\mu_{n} p_{n}\,
\left\{\, j_{\ell}\left( p_{n}r_{0}\right)\,n_{\ell}\left( p_{n}r_{0}\right)\,
-\, i\, j^{2}_{\ell}\left( p_{n}r_{0}\right)\,\right\}
\;\;\; .
\label{zeroD1channel}
\end{equation}
Here, $\mu_{n}$ and  $p_{n}$ are
the reduced mass and relative linear momentum of the two-meson system
at $E=E_{n}$, respectively.
Note that the imaginary part of $\Delta E_{n}$ is negative,
as it should be for resonance poles in the second Riemann sheet.
Below threshold we obtain poles on the real energy axis,
since $ij_{\ell}h^{(1)}_{\ell}$ becomes real for purely imaginary
arguments.
The latter poles represent two-meson bound states, as argued above.
For the following discussion we shall only consider poles above threshold.

For $\cal D$ we obtain
\begin{equation}
{\cal D}^{(\ell )}(E)\;\propto\;
\prod_{n}\,
\left( E-E_{n}-\Delta E_{n}\right)
\;\;\; .
\label{Dexpansion}
\end{equation}
Consequently, denoting the residue at the $n$-th pole
by $\alpha_{n}$, we get
\begin{equation}
\fndrs{0pt}{1}{-3pt}{{\cal D}^{(\ell )}(E)}\;\propto\;
\sum_{n}\,
\fnd{\alpha_{n}}{\left( E-E_{n}-\Delta E_{n}\right)}
\;\;\; ,
\label{DBWexpansion}
\end{equation}
which is nothing but a Breit-Wigner \cite{PR49p519} expansion
over a series of resonances, as employed in the isobar formalism
\cite{PLB84p349,PRD11p3165,PRD11p3183,PRD50p1972,NPA742p305}.

\subsection{Overlapping resonances}

Of course, things become more involved than in Eq.~(\ref{DBWexpansion})
when $\dlambda$ is not small and resonances start to overlap.
Overlapping resonances have been studied extensively in the past
\cite{NPA189p417}.
Here, it is no longer possible then to deduce simple approximations
for expression (\ref{D1channel}).

Besides extending the formalism of Ref.~\cite{PR88p1163}
to coupled channels and overlapping resonances,
our work also seems to interpolate between the results
of Ref.~\cite{HEPPH9710456} and Ref.~\cite{NPA744p127}.

\subsection{The \bm{K}-matrix}
The $K$-matrix,
which is related to the tangent(s) of the scattering phase shift(s),
is defined by
\begin{equation}
K\; =\; T\,\left[\, 1\, +\, iT\,\right]^{-1}
\;\;\; .
\label{kmatrix}
\end{equation}
As follows from the unitarity condition, $K$ is a real (symmetric) matrix
for real CM energy $E$.

In the one-channel approximation and in
a particular partial wave,
$K$ is given by the inverse of
expression~(\ref{cotangentd}) for the cotangent of
the scattering phase shift.
For more channels, relations like Eq.~(\ref{cotangentd})
become very complicated expressions in terms of
$\cal A$ and $\cal D$.
The reason is that the inverse of the expression (\ref{tmatrix})
has to be determined.
Numerically this is no problem, of course,
but analytically it is extremely tedious in the general multichannel case.
In particular, for a relation between
the common denominator $\cal D$ and $K$,
which is needed for the leading term in
expression (\ref{prodall}),
nothing simpel follows.
Moreover, the pole positions for both scattering and production
stem from $\cal D$, and not from $K$.
Hence, the excercise to express the production amplitude
in terms of the $K$-matrix seems pointless.

\section{Summary and Concluding Remarks}

The two-meson production amplitude~(\ref{prodall})
has been rigorously calculated, to all orders,
from a relatively general expression for
a two-meson scattering amplitude (Eq.~(\ref{tmatrix})) dominated by
$s$-channel resonances. The latter had already been succesfully tested
for $c\bar{c}$ and $b\bar{b}$ states, mesons with open charm and
bottom, and also in the light-quark sector.

One might object that a model with no $t$-channel exchanges is too
restricted for drawing general conclusions. However, one should
be aware of the --- quoting T\"{o}rnqvist \cite{PRL77p2333} ---
``well-known dual-model result for $\bar{q}q$ resonances that a sum
of $s$-channel resonances also describes $t$- and $u$-channel
phenomena.'' In the context of duality, Harari \cite{PRL26p1400}
formulated a necessary condition for an $s$-channel description to
reproduce certain $t$-channel effects, namely the existence of ``strong
correlations between the different $s$-channel resonances.'' Well, this
is exactly what our infinite RSE sum over confinement states guarantees.
Further proof showing the RSE model to be realistic is its correct
threshold behaviour in elastic $\pi\pi$ scattering \cite{HEPPH0702117}.

Another possible critique of our method could be its NR
nature. Nevertheless, in practical phenomenological applications to
spectroscopy and elastic scattering, relative momenta and reduced masses
in the two-meson channels have been consistently defined in a relativistic
way, thus ensuring proper kinematics at much higher energies than the
underlying NR formalism seems to support. Such a minimal treatment of
relativity is indeed common practice in many relativised quark models. Our
successful description of the spectroscopy and scattering properties of the
light scalar mesons \cite{PLB641p265} provides additional evidence that
this approach to relativity is reasonable. This is also supported by our
very recent first application of the present production formalism in the
single-channel case \cite{JPG34p1789}.

It thus seems fair to conclude that production amplitudes
can in general contain terms which are not
proportional to scattering $T$-matrix elements
and, moreover, that the proportionality coefficients are complex.

\section*{Acknowledgements}

We wish to thank I.~J.~R. Aitchison, D.~V.~Bugg and C.~Hanhart for
useful discussions.
This work was supported in part by the {\it Funda\c{c}\~{a}o para a
Ci\^{e}ncia e a Tecnologia} \/of the {\it Minist\'{e}rio da Ci\^{e}ncia,
Tecnologia e Ensino Superior} \/of Portugal, under contract
PDCT/\-FP/\-63907/\-2005.

\appendix

\section{Generic relation between production and scattering}
\label{PvsS}

In order to arrive at a relation equivalent to Eq.~(\ref{ImaTstera})
for the here proposed scattering and production amplitudes,
we define
\begin{equation}
T^{(\ell )}_{ij}\, =\,
-\, 2\sqrt{\mu_{i}p_{i}\mu_{j}p_{j}\,}\,
t_{\ell}(i\to j)\, =\,
-\, 2\dlambda^{2}\,
\sqrt{\mu_{i}p_{i}\mu_{j}p_{j}\,}\,
j_{\ell}\left( p_{i}r_{0}\right)\, j_{\ell}\left( p_{j}r_{0}\right)\;
\fnd{{\cal A\,}^{(\ell)}_{ij}}{{\cal D\,}^{(\ell)}}
\;\;\; .
\label{Tpartialwave}
\end{equation}
For this object,
also using relations (\ref{UnitaircompleteTell}),
one easily finds
\begin{eqnarray}
& &
\;\;\;\;
\sum_{\nu}\,
T^{(\ell )\ast}_{i\nu}\,
T^{(\ell )}_{\nu j}\, =\;
4\dlambda^{4}\,
\sqrt{\mu_{i}p_{i}\mu_{j}p_{j}\,}\,
j_{\ell}\left( p_{i}r_{0}\right)\,
j_{\ell}\left( p_{j}r_{0}\right)\,
\sum_{\nu}\,
\mu_{\nu}p_{\nu}\,
j^{2}_{\ell}\left( p_{\nu}r_{0}\right)\,
\fnd{
{\cal A\,}^{(\ell)\ast}_{i\nu}\,
{\cal A\,}^{(\ell)}_{\nu j}}
{\abs{{\cal D\,}^{(\ell)}}^{2}}
\label{Tunitair}\\ [10pt] & &
\!\!\!\!\!\! =\,
\fnd{\dlambda^{2}}{i}\,
\sqrt{\mu_{i}p_{i}\mu_{j}p_{j}\,}\,
j_{\ell}\left( p_{i}r_{0}\right)\,
j_{\ell}\left( p_{j}r_{0}\right)\,
\left\{\,
\fnd{{\cal A\,}^{(\ell)\ast}_{ij}}{{\cal D\,}^{(\ell)\ast}}
\, -\,
\fnd{{\cal A\,}^{(\ell)}_{ij}}{{\cal D\,}^{(\ell)}}
\,\right\}
\, =\,
\fnd{1}{2i}\left\{\,
T^{(\ell )}_{ij}\,
-\,
T^{(\ell )\ast}_{ij}
\,\right\}
\, =\,
\Imag{T^{(\ell )}_{ij}}
\; .
\nonumber
\end{eqnarray}

Furthermore, we define
\begin{equation}
A^{(\ell )}_{\alpha i}\; =\;
\sqrt{\mu_{i}\, p_{i}\,}
j_{\ell}\left( p_{i}r_{0}\right)\,
\left\{ g_{\alpha i}
\, -\,
2i\dlambda^{2}\;
\sum_{\nu}\,
\mu_{\nu}\, p_{\nu}\,
j_{\ell}\left( p_{\nu}r_{0}\right)\,
h^{(1)}_{\ell}\left( p_{\nu}r_{0}\right)\,
g_{\alpha\nu}\,
\fnd{{\cal A\,}^{(\ell )}_{i\nu}}{{\cal D\,}^{(\ell )}}\,
\right\}
\;\;\; ,
\label{PAmpdef1}
\end{equation}
for which,
by substituting definition (\ref{Tpartialwave}),
we may also write
\begin{equation}
A^{(\ell )}_{\alpha i}\; =\;
g_{\alpha i}\,
j_{\ell}\left( p_{i}r_{0}\right)\,
\sqrt{\mu_{i}\, p_{i}\,}
\, +\,
i\,\sum_{\nu}\,
g_{\alpha\nu}\,
\sqrt{\mu_{\nu}\, p_{\nu}\,}\,
h^{(1)}_{\ell}\left( p_{\nu}r_{0}\right)\,
T^{(\ell )}_{i\nu}
\;\;\; .
\label{PAmpdef}
\end{equation}

For this object we study, in accordance with relation (\ref{ImaTstera}),
the imaginary part
\begin{eqnarray}
\lefteqn{\Imag{A^{(\ell)}_{\alpha i}}\,
=\,
\,\sum_{\nu}\,
g_{\alpha\nu}\,
\sqrt{\mu_{\nu}\, p_{\nu}\,}\,
\frac{1}{2i}\,
\left\{
ih^{(1)}_{\ell}\left( p_{\nu}r_{0}\right)\,
T^{(\ell)}_{i\nu}
+\,
ih^{(2)}_{\ell}\left( p_{\nu}r_{0}\right)\,
{T^{(\ell)}_{i\nu}}^{\ast}
\right\}
}
\label{Impartialamplitude1}\\ [10pt] & & =\,
\frac{1}{2}\,\sum_{\nu}\,
g_{\alpha\nu}\,
\sqrt{\mu_{\nu}\, p_{\nu}\,}\,
\left\{
j_{\ell}\left( p_{\nu}r_{0}\right)\,
\left( T^{(\ell)}_{i\nu}+{T^{(\ell)}_{i\nu}}^{\ast}\right)
\, +\,
in_{\ell}\left( p_{\nu}r_{0}\right)\,
\left( T^{(\ell)}_{i\nu}-{T^{(\ell)}_{i\nu}}^{\ast}\right)
\right\}
\nonumber\\ [10pt] & & =\,
\sum_{\nu}\,
g_{\alpha\nu}\,
\sqrt{\mu_{\nu}\, p_{\nu}\,}\,
\left\{
j_{\ell}\left( p_{\nu}r_{0}\right)\,
\Real{T^{(\ell)}_{i\nu}}
\, -\,
n_{\ell}\left( p_{\nu}r_{0}\right)\,
\Imag{T^{(\ell)}_{i\nu}}
\right\}
\;\;\;,
\nonumber
\end{eqnarray}
where we denote
the spherical Hankel function of the second kind
by $h^{(2)}_{\ell}=h^{(1)\ast}_{\ell}=j_{\ell}-in_{\ell}$.

Next, we use the fact that $\Real{T}=T^{\ast}+i\Imag{T}$,
and, moreover, substitute subsequently relations
(\ref{Tunitair}) and (\ref{PAmpdef}):
\begin{eqnarray}
\lefteqn{\Imag{A^{(\ell)}_{\alpha i}}\,
=\,
\sum_{\nu}\,
g_{\alpha \nu}\,
\sqrt{\mu_{\nu}\, p_{\nu}\,}\,
\left\{
j_{\ell}\left( p_{\nu}r_{0}\right)\,
{T^{(\ell)}_{i\nu}}^{\ast}
\, +\,
i\,
h^{(1)}_{\ell}\left( p_{\nu}r_{0}\right)\,
\Imag{T^{(\ell)}_{i\nu}}
\right\}}
\label{Impartialamplitude}\\ [10pt] & & =\,
\sum_{\nu}\,
g_{\alpha \nu}\,
\sqrt{\mu_{\nu}\, p_{\nu}\,}\,
j_{\ell}\left( p_{\nu}r_{0}\right)\,
{T^{(\ell)}_{i\nu}}^{\ast}
\, +\,
i\,
\sum_{\nu '}\,
\sum_{\nu}\,
g_{\alpha\nu}\,
\sqrt{\mu_{\nu}\, p_{\nu}\,}\,
h^{(1)}_{\ell}\left( p_{\nu}r_{0}\right)\,
T^{(\ell)}_{\nu '\nu}\,
{T^{(\ell)}_{i\nu '}}^{\ast}
\nonumber\\ [10pt] & & =\,
\sum_{\nu}\,
{T^{(\ell)}_{i\nu}}^{\ast}\,
\left\{
g_{\alpha \nu}\,
j_{\ell}\left( p_{\nu}r_{0}\right)\,
\sqrt{\mu_{\nu}\, p_{\nu}\,}\,
\, +\,
i\,
\sum_{\nu '}\,
g_{\alpha\nu '}\,
\sqrt{\mu_{\nu '}\, p_{\nu '}\,}\,
h^{(1)}_{\ell}\left( p_{\nu '}r_{0}\right)\,
T^{(\ell)}_{\nu\nu '}
\right\}
\nonumber\\ [10pt] & & =\,
\sum_{\nu}\,
{T^{(\ell )}_{i\nu}}^{\ast}\, A^{(\ell )}_{\alpha\nu}\;
\; .
\nonumber
\end{eqnarray}
This demonstrates that for our amplitudes a relation exists
which is equivalent to the one shown in Eq.~(\ref{ImaTstera}).

\newcommand{\pubprt}[4]{{#1 {\bf #2}, #3 (#4)}}
\newcommand{\ertbid}[4]{[Erratum-ibid.~{#1 {\bf #2}, #3 (#4)}]}
\def\AIPCP{AIP Conf.\ Proc.}
\def\DAP{Annalen Phys.}
\def\EPJA{Eur.\ Phys.\ J.\ A}
\def\EPJC{Eur.\ Phys.\ J.\ C}
\def\IJMPA{Int.\ J.\ Mod.\ Phys.\ A}
\def\IJTPGTNO{Int.\ J.\ Theor.\ Phys.\ Group Theor.\ Nonlin.\ Opt.}
\def\JPCS{J.\ Phys.\ Conf.\ Ser.}
\def\JPG{J.\ Phys.\ G}
\def\NCA{Nuovo Cim.\ A}
\def\NPA{Nucl.\ Phys.\ A}
\def\NPB{Nucl.\ Phys.\ B}
\def\PAN{Phys.\ Atom.\ Nucl.}
\def\PLB{Phys.\ Lett.\ B}
\def\PR{Phys.\ Rev.}
\def\PRD{Phys.\ Rev.\ D}
\def\PRL{Phys.\ Rev.\ Lett.}
\def\PTP{Prog.\ Theor.\ Phys.}
\def\ZPC{Z.\ Phys.\ C}

\end{document}